\documentstyle[sprocl,epsfig,pstricks]{article}
\newcommand{\ra}{\rightarrow}

\def\ra{\rightarrow}

\def\be{\begin{equation}}
\def\ee{\end{equation}}
\def\bea{\begin{eqnarray}}
\def\eea{\end{eqnarray}}

\def\gsim{\,\lower.25ex\hbox{$\scriptstyle\sim$}\kern-1.30ex%
\raise 0.55ex\hbox{$\scriptstyle >$}\,}
\def\lsim{\,\lower.25ex\hbox{$\scriptstyle\sim$}\kern-1.30ex%
\raise 0.55ex\hbox{$\scriptstyle <$}\,}

\newcommand{\ccb}{c\bar{c}}

\begin{document}

\title{HEAVY QUARK AND JET PRODUCTION \\ BY REAL AND VIRTUAL PHOTONS
\footnote{Invited talk given at the XVIIth International Conference
          on Physics in Collision, Bristol, UK, June 25-27, 1997} 
}

\author{ F. SEFKOW}

\address{Physik-Institut der Universit\"at Z\"urich \\
         CH-8057 Z\"urich, Switzerland}


\maketitle\abstracts{
Recent results from the H1 and ZEUS collaborations 
on hard QCD processes in $ep$ interactions are reviewed.
The topics cover jet shapes, the structure of real and virtual photons, 
and the production of $J/\Psi$ mesons and open charm.}

\section{Introduction}

Heavy flavours and jets 
in the hadronic final state 
carry much information about the dynamics 
of the unobservable quarks and gluons. 
They thus provide access to the underlying hard partonic processes 
of high energy $ep$ interactions.
The aim is to test the description of the production processes 
by perturbative QCD, 
and to extract information
on the parton distributions of the initial state hadron.  

At HERA, electrons (or positrons) of 27.5 GeV
collide with 820 GeV protons,
the center-of-mass energy is $\sqrt{s}=$ 300 GeV.
The bulk of the scattering reactions occur in the photoproduction regime
where the momentum transfer 
$Q^2 \approx 0$, 
i.e. the exchanged photon is almost real;
data in the Deep Inelastic Scattering (DIS) regime correspond to large
$\gamma$ virtualities $Q^2>>1$ GeV$^2/c^2$.    
In both regimes, however, the $\gamma p$ center-of-mass energies $W$ 
are 
typically as high as $\approx$ 100 - 200 GeV 
and open up the phase space for hard QCD processes. 
The HERA experiments H1 and ZEUS 
thus extend the range of fixed-target photoproduction 
experiments by an order of magnitude. 

This review begins with results from the study of jets 
in photoproduction and DIS, 
the focus is on the structure of both real and virtual photons.
In the second part, the production of bound and open charm states
will be discussed in terms of perturbative QCD models.

\section{Jets in Photoproduction}

In Leading Order QCD, a photon can either interact directly 
with a parton in the proton (``direct'' photoproduction,
$\gamma q\ra qg$ and $\gamma g\ra q\bar{q}$),
or it can fluctuate into a 
quark-antiquark pair,
which may form a vector meson state or may develop 
a quark-gluon structure without forming a hadronic bound state
(``anomalous '' characteristics of the photon). 
A parton from that state
interacts with a parton from the proton (``resolved'' photoproduction;
most important are 
$qq' \ra qq'$, $qg \ra qg$ and $gg \ra gg$).
Since the parton carries only a fraction $x_{\gamma}$ of the 
photon's momentum, the ``resolved'' events at HERA exhibit 
a stronger boost into the proton (``forward'') direction,
towards positive pseudo-rapidities $\eta$.%
\footnote{The pseudo-rapidity is defined as 
$\eta = - \ln \tan (\theta /2)$, where $\theta$ is
the polar angle measured with respect to  
the $z$ axis, taken to be along the direction 
of the incident proton beam.}

\subsection{Jet Shapes}

A necessary prerequisite for 
the interpretation of jet photoproduction data in partonic terms is
being able to model the internal structure of hadron jets.  
Let the jet be defined by a cone algorithm
which combines hadrons within a cone of radius $R$,
measured in the $(\eta ,\phi)$ plane.
A suitable quantity to describe the jet shape is $\Psi (r)$,
defined as the fraction of the jet's transverse energy contained 
in a sub-cone of radius $r$. 
$\Psi (r)$ increases towards 1 with $r$ approaching $R$; 
the more collimated the jet, 
the steeper the rise and the higher $\Psi$ at fixed $r$.

\begin{figure}[tb]\centering
\epsfig{file=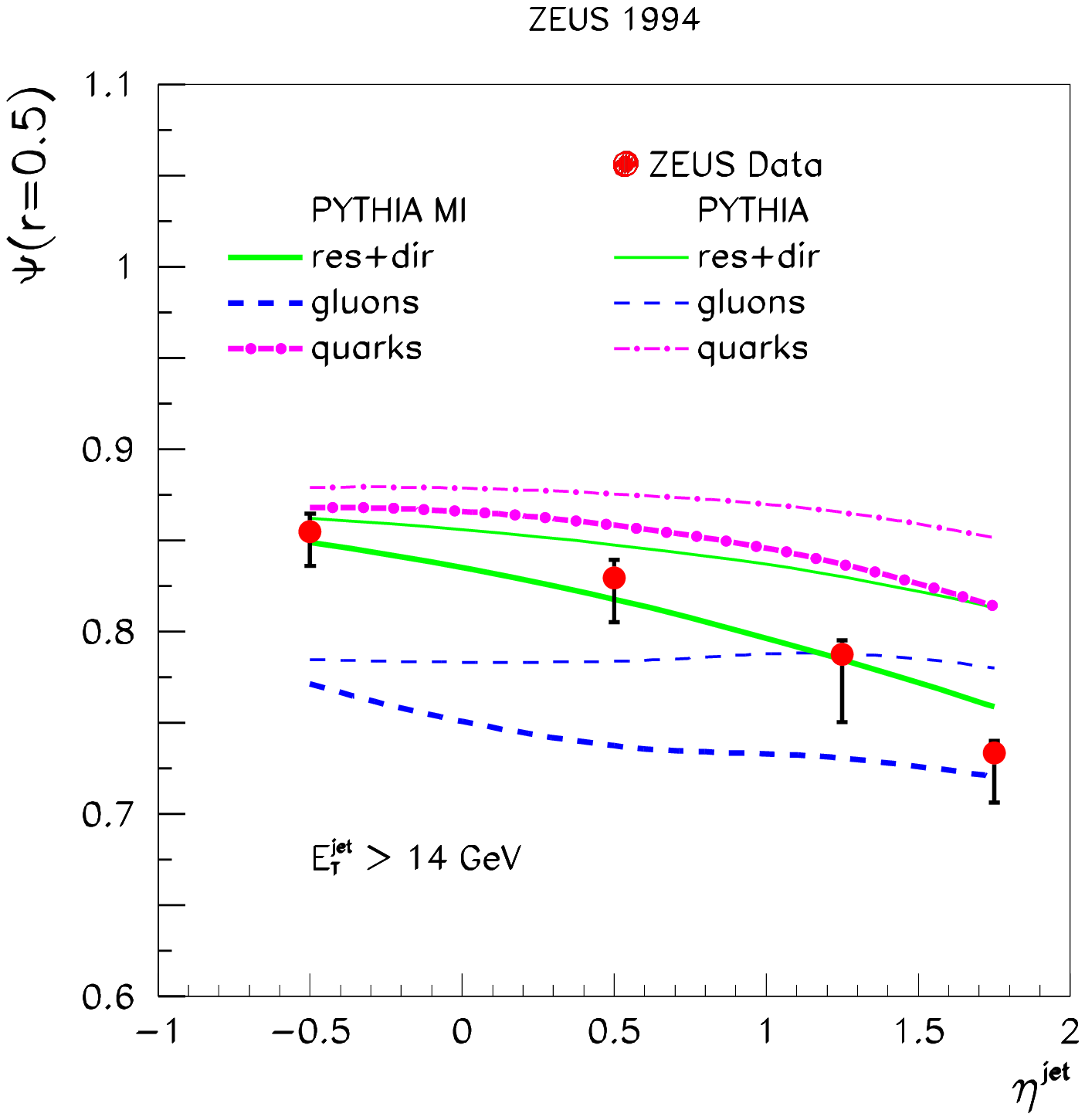,width=6.1cm}
\epsfig{file=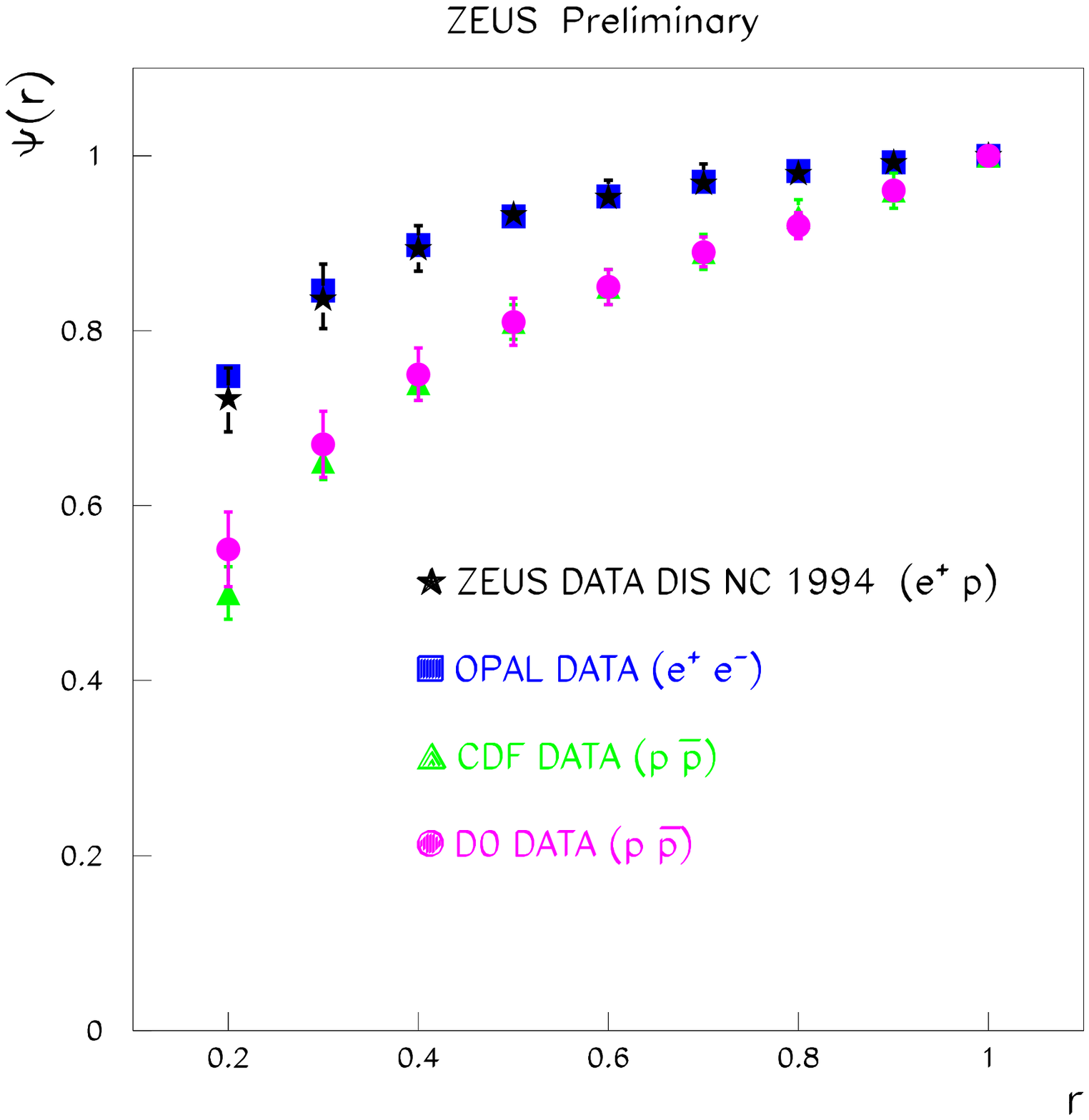,width=5.6cm}

\unitlength1cm
\begin{picture}(0,0)
\put(-1,1.5){(a)}
\put(5,1.5){(b)}
\end{picture}
\vspace{-0.5cm}
\caption[dummy]{\label{fig:jetshape}
        Jet shape (fraction of transverse energy in a sub-cone of 
        radius $r$) for a jet defined by a cone of radius $R=1$, 
        (a) in photoproduction, at fixed  
            $r=0.5$ as function of rapidity,
        (b) in DIS at HERA, as function of $r$, compared with 
            LEP and Tevatron data.  
} 
\end{figure}
In Fig.~\ref{fig:jetshape}a $\Psi (r=0.5)$ 
as measured by ZEUS~\cite{zeusjetsh}
is displayed as a function of the jet pseudo-rapidity $\eta$,
for jets with transverse energy $E_{\perp}>$ 14 GeV and $R=1$.
The result clearly shows that the jets are broader in the forward region. 
Predictions of the 
QCD Monte Carlo program PYTHIA~\cite{pythia}
are shown for comparison. 
The thick curve 
which gives the best description of the data. 
has been obtained with
the inclusion of multi-parton interactions in ``resolved'' processes.
The predictions are also shown separately for 
jets initiated by either quarks or gluons.
Gluon jets are broader since they have higher color charge 
so that QCD radiation is more likely to occur.
In the generator program, the average broadening with $\eta$ 
can be traced to an increased fraction of gluon-initiated jets, 
which in turn is due to the increased contribution of ``resolved'' events 
in the forward direction. 

It is interesting to note that in DIS, at high $Q^2>100$ GeV$^2/c^2$,
the jet shapes measured at HERA~\cite{zeusjetsh} agree very well with those 
observed at LEP~\cite{jetsh_LEP}
(Fig.~\ref{fig:jetshape}b). 
This is expected in the quark parton model, where a quark is ``kicked'' 
out of the proton and produces the jet. 
The function $\Psi (r)$ 
exhibits a distinctly steeper rise 
than that observed in $\bar{p}p$ interactions~\cite{jetsh_pp}, 
where predominantly gluon initiated jets are produced.

Jet shapes in photoproduction have also been calculated 
analytically using  Next to Leading Order (NLO) 
QCD matrix elements~\cite{klasenjetsh}.  
In order to emulate effects of the experimental jet finding procedures, 
a parameter $R_{sep}$ has to be introduced into  
the theoretical calculation;
$R_{sep}$ specifies the maximum distance of two partons 
to be merged into one jet. 
Given the freedom in this jet finding parameter,
the measured jet shapes can be reproduced.

\subsection{Dijet angular distribution}
\label{sec:dijetgp}

In the Leading Order (LO) picture, the kinematics of the partonic $2 \ra 2$ 
process can be fully reconstructed in dijet events 
from the jets' energies and angles. 
Such an analysis can reveal dynamical features of 
the strong interactions mediating the jet production.

One defines an observable related to 
the momentum fraction of the parton in the photon,
$x_{\gamma}^{OBS} = \sum _{jets} E_{\perp}^{jet} \exp (-\eta^{jet}) 
/ 2yE_e$    
where $E_e$ is the $e$ beam energy,
and $y=1-E_{\gamma}/E_e$ 
can be obtained from the scattered electron 
or from all final state hadrons. 
For the scattering angle in the partonic center-of-mass system one has
$\cos \theta ^* = \tanh (\eta^{jet\,2}-\eta^{jet\,1})/2$.
Although beyond LO only the sum of ``direct'' and ``resolved''
processes can be unambiguously defined, and the $2\ra 2$ kinematics formulae 
do not apply exactly anymore, 
these quantities -- being defined as hadronic observables --  
are meaningful at any order.

\begin{figure}[tbp]\centering
\epsfig{file=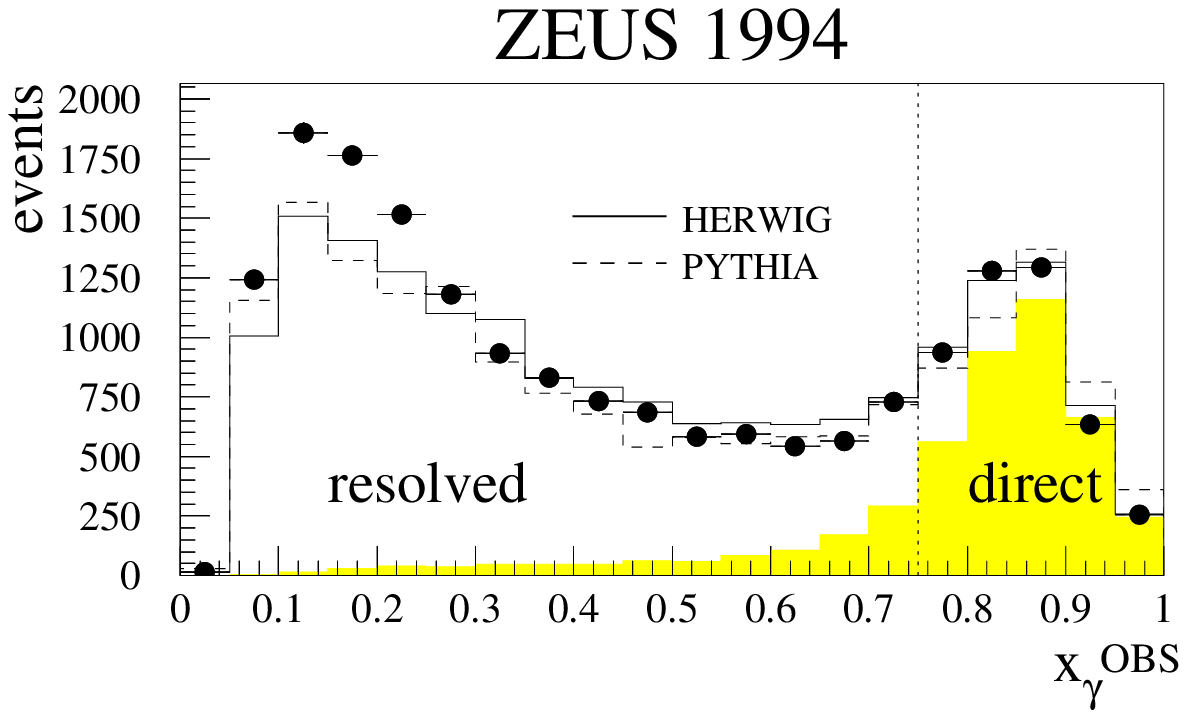,width=5.8cm}
\epsfig{file=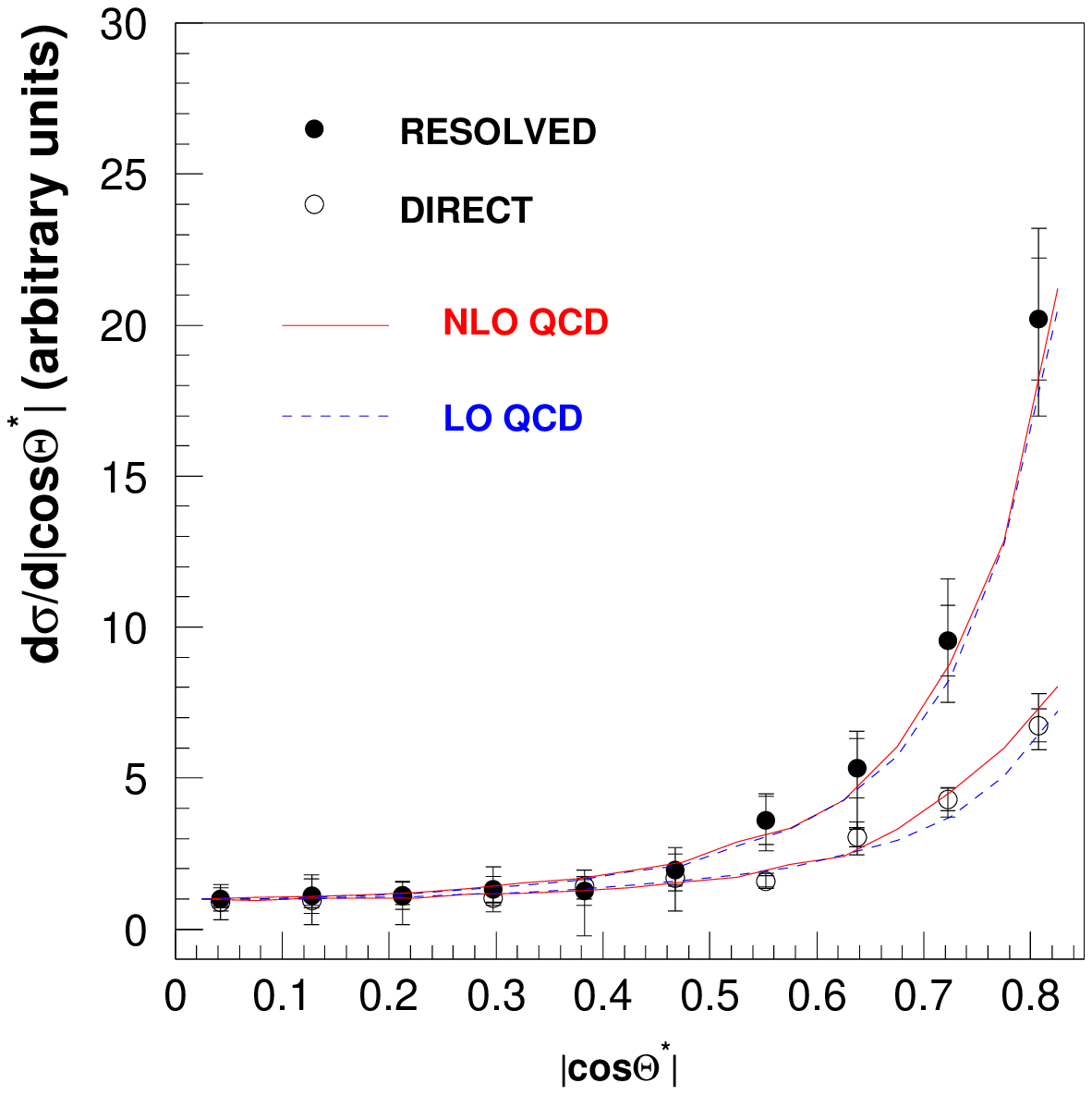,width=5.8cm}

\unitlength1cm
\begin{picture}(0,0)
\put(-1,3){(a)}
\put(5,5.7){(b)}
\end{picture}
\vspace{-0.8cm}
\caption[dummy]{\label{fig:dijet}
        (a) Uncorrected $x_{\gamma}^{OBS}$ distribution
            for dijet events. 
        (b) Dijet angular distribution in the parton parton 
            center-of-mass system.
} 
\end{figure}

The uncorrected $x_{\gamma}$ distribution from ZEUS~\cite{zeusdijetang}
is shown in Fig.~\ref{fig:dijet}a.
It is qualitatively reproduced by the Monte Carlo calculations
shown in the same figure.
For direct processes, $x_{\gamma}^{OBS}$ is expected to be close to 1. 
The data is thus split into a ``direct'' sample ($x_{\gamma}>0.75$)
and a ``resolved'' sample ($x_{\gamma}<0.75$). 

In QCD, the dominant Leading Order diagrams for direct processes 
involve quark (spin $\frac{1}{2}$) exchange, 
whereas for the ``resolved'' case, 
gluon exchange (spin 1) dominates. 
The angular distributions are thus predicted to be different: 
$\sim |1-\cos \theta ^*|^{-1}$ in the first case, 
and a steeper $|1-\cos \theta ^*|^{-2}$ behavior in the latter. 
This expectation is nicely confirmed 
by the ZEUS measurements~\cite{zeusdijetang}
shown in Fig~\ref{fig:dijet}b. 
The distributions reflect the spin of the exchange
as well as the relative contributions of direct and resolved processes
to each sample.  
Moreover, NLO calculations~\cite{harrisowensjets}
exhibit the same pattern and also agree well with the data.

\subsection{Photon Structure}

The jet cross sections evidently depend on the parton distributions 
in the photon.
At HERA, the proton acts as hadronic probe,
and the data are sensitive to both the quark and the gluon
content of the photon. 
The measurements have provided the 
first determination of the gluon momentum distribution
by H1~\cite{h1gxgamma}.
The recent analysis presented here~\cite{h1fxgeff} investigates  
the scale dependence of the parton distribution 
and confronts it to QCD predictions.

H1 has measured the double differential dijet cross section as 
a function of $x_{\gamma}^{OBS}$ and $E_T^2$, 
the second highest transverse jet energy squared, 
in the range $0.1<x_{\gamma}<1$ and $E_T^2>8$ GeV$^2$.
The data have been analyzed in terms of an effective 
parton distribution -- i.e.\, not differentiating quarks and following the concept of Ref.~\cite{combmaxwell}.
This is possible since the most important leading order matrix elements 
in ``resolved'' photoproduction have angular distributions similar 
in shape, 
they mainly differ in overall magnitude by the ratio of color factors. 
Consequently, the cross section can be described by means of a 
single effective subprocess with matrix element $|M_{SES}|^2$. 
One defines {\em effective} parton distributions for photon 
{\em and} proton:
\be
f_{eff}(x,p_{\perp}^2) = \sum_{flavours}
(q(x,p_{\perp}^2) + \bar{q}(x,p_{\perp}^2))
+ \frac{9}{4}g(x,p_{\perp}^2) \;\; ,
\ee
where $q$ and $g$ denote the quark and gluon densities, respectively,
$x$ the momentum fraction and $p_{\perp}^2$ the factorization scale of the
process given by the parton's transverse momentum squared.
The dijet cross section then factorizes to good approximation, symbolically:
\be
\sigma  \sim f_{eff}^{\gamma}(x_{\gamma},p_{\perp}^2) 
       \cdot f_{eff}^p(x_p,p_{\perp}^2) \cdot |M_{SES}|^2 \;\; .
\ee    

\begin{figure}[bt]\centering
\epsfig{file=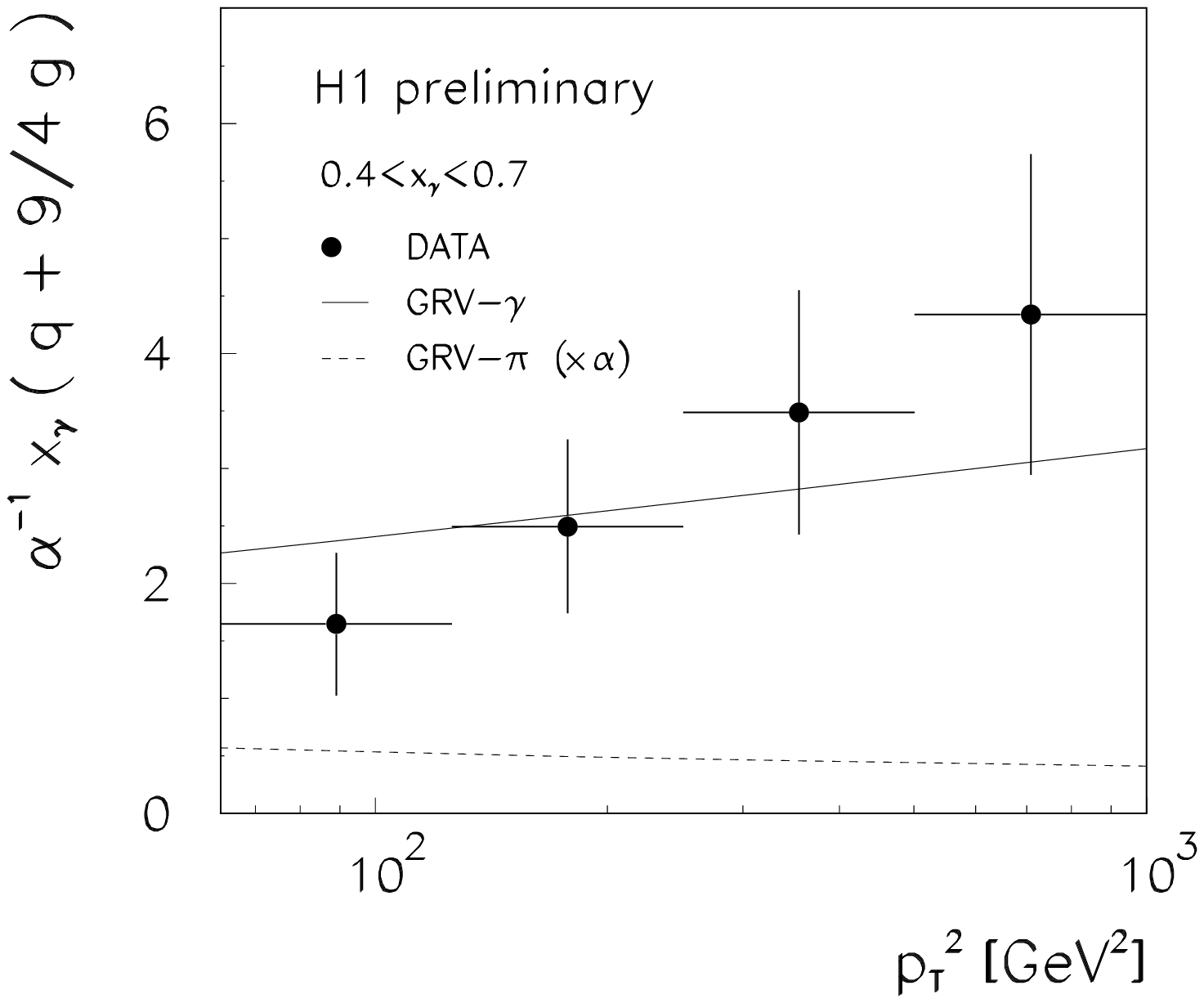,width=5.7cm}
\epsfig{file=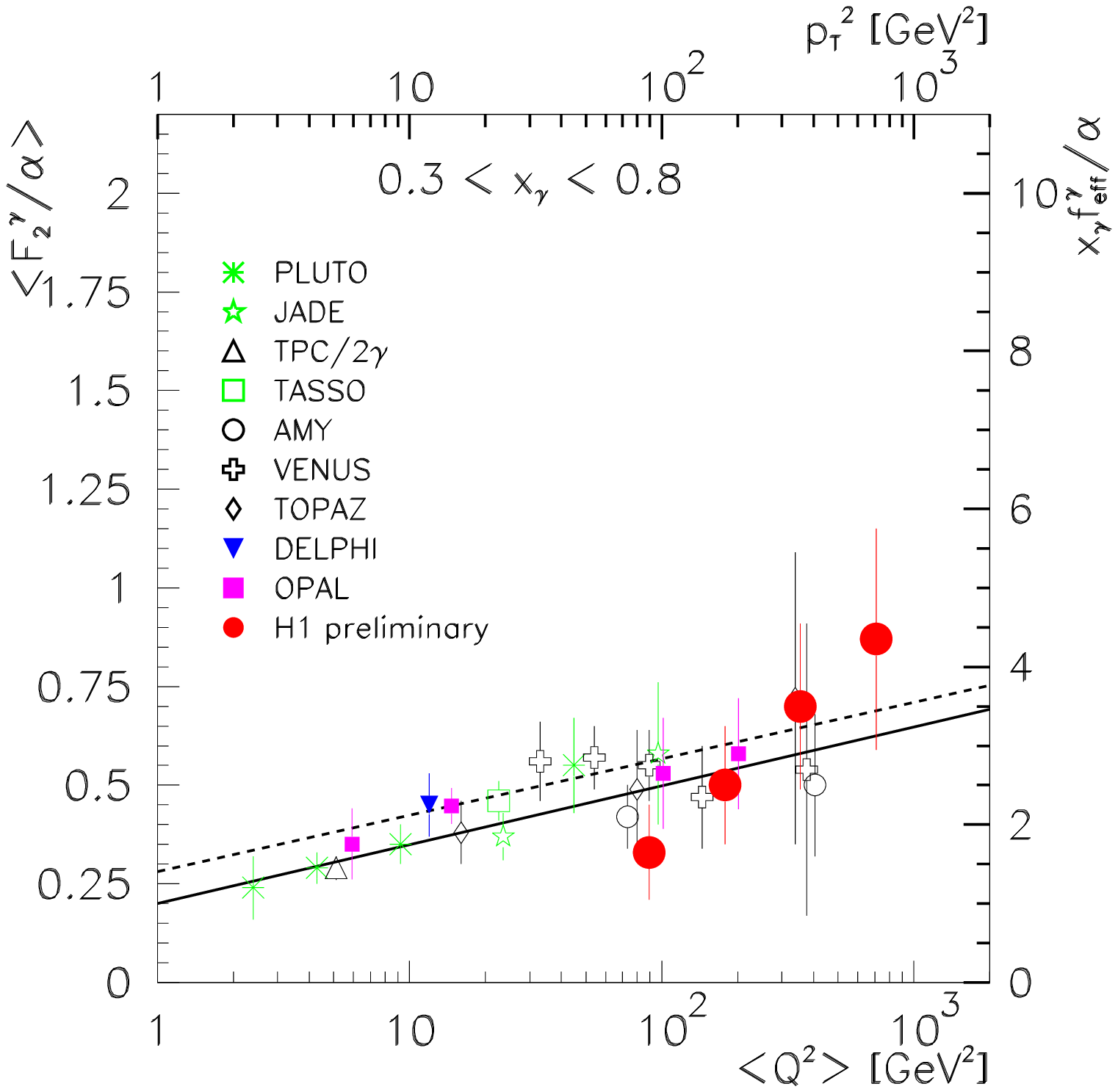,width=5.7cm}

\unitlength1cm
\begin{picture}(0,0)
\put(-1.1,4.6){(a)}
\put(4.4,4.6){(b)}
\end{picture}
\vspace{-0.7cm}
\caption[dummy]{\label{fig:effef}
        (a) Scale dependence of the effective parton density 
            $f_{eff}^{\gamma}$ of the photon.
        (b) Dependence of the photon structure function $F_2^{\gamma}$,
            measured in two photon reactions,  
            on the scale given by the virtuality $Q^2$ of the probing
            photon. $f_{eff}^{\gamma}$ was divided by 5 (see text) 
            and is shown as function 
            of the scale given by the parton transverse momenta 
            $p_t^2$~\cite{rickphd}.
            The lines represent calculations in the FKP model~\cite{fkp}
            with different parameter settings.
} 
\end{figure}

The effective parton distribution $f_{eff}^{\gamma}$ extracted from the 
double differential cross section is shown in Fig.~\ref{fig:effef}a
as function of the scale $p_{\perp}^2$ for the range
$0.4<x_{\gamma}<0.7$. 
The data clearly show a direct dependence on the scale;  
This positive scaling violation
is remarkably different from the behavior of hadronic structure 
functions,
where the dependence on the scale enters only via the 
QCD evolution equations and,  
at large momentum fractions, is negative. 
For comparison 
the GRV parameterization~\cite{grv_gamma} of the photon structure 
is shown (full line).
In a vector meson dominance picture the effective parton density
would be proportional to a vector meson structure function; 
here the effective parton density of the pion is shown, 
scaled by the probability of the photon fluctuating into a vector meson. 
this is also shown for comparison as the dashed line.   
The behavior of the photon is a genuine QCD prediction. 
It originates from the inhomogeneous term in the Altarelli-Parisi
evolution equations which is only present for the photon, 
due to its point-like coupling to quarks, 
and is referred to as its ``anomalous component''. 

The H1 result ties in nicely with the scaling behavior of the photon 
structure function $F_2^{\gamma}$ as determined in two-photon experiments,
where a highly virtual photon probes the quark content of a real photon. 
QCD predicts the scaling to be $F_2^{\gamma}\sim \ln Q^2/\Lambda^2_{QCD}$.
A similar dependence can be expected for $f_{eff}^{\gamma}$
at large $x_{\gamma}$, 
since here the gluon contribution to it is
small according to the H1 measurement~\cite{h1gxgamma}.  
From the GRV parameterization of quark and gluon distributions,
the ratio $f_{eff}^{\gamma}/F_2^{\gamma}$ is  
expected to be about 5 and approximately constant.  
Fig.~\ref{fig:effef}b~\cite{rickphd} 
shows the H1 result, scaled by this factor, 
overlaid onto a compilation of 
$F_2^{\gamma}$ results from $e^+e^-$ experiments~\cite{f2gammacompi}.
The HERA data are comparable in precision, and they 
extend the kinematic range towards a maximum scale of 
$p_{\perp}^2=1250\,$GeV$^2/c^2$.

\section{Jets in Deep Inelastic Scattering}

In the previous section, it was discussed how 
the structure of the photon
can be ``resolved''
by partons inside the proton. 
The ``resolution'' is given by the hard scale of the partonic
sub-process, e.g.\ the transverse energy squared, $p_{\perp}^2$. 
On the other hand, deep inelastic $ep$ scattering (DIS) is usually described 
in the quark parton model, where a virtual, point-like photon 
probes the structure of the proton, the ``resolution''
being given by the negative 4-momentum transfer $Q^2$. 
In the kinematic regime of 
DIS, where $Q^2$ is low, but there is another hard scale 
$p_{\perp}^2 > Q^2$ present,     
it is an open question whether this is still an 
adequate description.
HERA offers the possibility to probe this interesting transition region
between DIS and the photoproduction limit. 

\subsection{Dijet Rates}

2+1 jet (dijet) events%
\footnote{The proton remnant jet is counted separately as ``+1''.}
in QCD are due to ${\cal O}(\alpha_s)$ 
processes, namely radiation of a hard gluon off the scattered 
quark (QCD-Compton), or boson gluon fusion (BGF). 
At low $Q^2<100$ GeV$^2/c^2$,
BGF is strongly dominant. 

\begin{figure}[bt]\centering
\epsfig{file=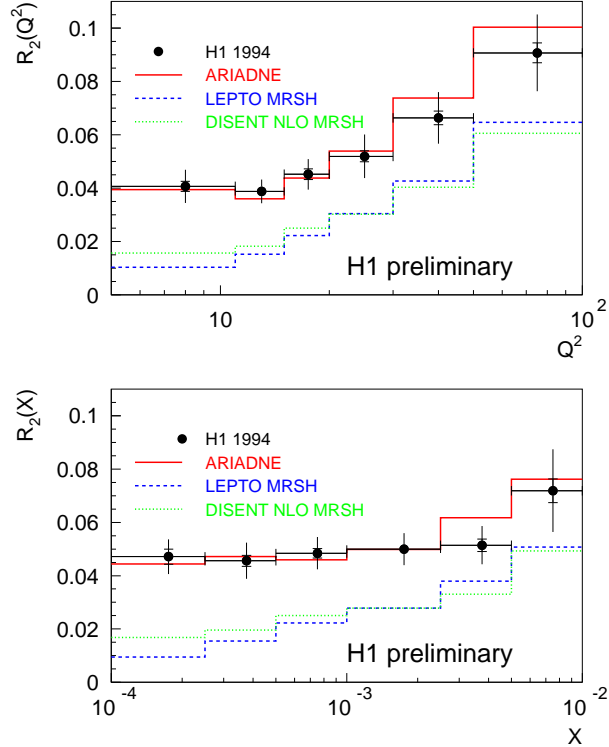,width=8cm}
\vspace{-0.4cm}
\caption[dummy]{\label{fig:r2}
        Dijet rate in DIS as function of $Q^2$ and $x_{Bj}$, 
        compared to ARIADNE (full histogram),
        LEPTO (dashed) and DISENT (light dashed) calculations.
} 
\end{figure}

The dijet rate $R_2$ measured by H1~\cite{spiekermoriond} is shown in 
Fig.~\ref{fig:r2} as a function of $Q^2$ 
and the Bjorken scaling variable $x_{Bj}$.
The jets were identified 
in the hadronic center-of-mass system 
by a cone algorithm with $R=1$ 
and were required to have $E_T^*>5$ GeV. 
Various predictions are compared to 
the data. 
The Color Dipole Model (CDM) 
as implemented in the ARIADNE~\cite{ariadne} Monte Carlo
program describes the data well.
The LEPTO~\cite{lepto} Monte Carlo
contains the exact LO (${\cal O}(\alpha_s)$) matrix elements 
and parton showers
in leading log approximation (MEPS).  
It falls short in describing the measured rate, 
in particular at lower $Q^2$. 
The same holds for 
a full NLO QCD calculation (DISENT~\cite{disent}) 
where the jet algorithm is used to find jets from the final state partons. 
This discrepancy is unlikely to be an effect 
of missing hadronization effects, since these were found to be below
20\% in the two Monte Carlo models. 

The failure of the NLO calculation to reproduce the dijet rate may be 
an indication that higher order corrections are important in the 
kinematic region probed here. 
The CDM 
also predicts additional energy flow (hadronic activity) 
in the photon fragmentation region of the $\gamma p$ center-of-mass system, 
increasing as $Q^2$ decreases,
whereas in the parton shower model,  
this quantity remains small and constant.  
The $Q^2$ dependence of the H1 data is reproduced by 
the CDM prediction~\cite{spiekermoriond}. 
It is possible to introduce additional processes with increased activity
in this region into the conventional QCD models by allowing 
or a partonic structure of the virtual photon~\cite{sasgam,dreesgodbole}, 
analogous to the description of jet production by real photons. 
This will be investigated further in the future.  

\subsection{Virtual Photon Structure}

The parton densities in the virtual photon
are expected in QCD to be suppressed with increasing virtuality $Q^2$;
for $Q^2>p_t^2$, the point-like component dominates.
For example, according to the phenomenological ansatz
of Drees and Godbole~\cite{dreesgodbole}, 
following the analysis of Borzumati and Schuler~\cite{borzumati},
the $Q^2$-dependent suppression with respect to the 
densities in the real photon is parameterized as 
\be
f_{q|\gamma ^*}(x, p_t^2, Q^2) = f_{q|\gamma }(x, p_t^2) \cdot L
\;\;\;\mbox{with}\;\;\; L =
\frac{\ln [(p_t^2+\omega ^2)/(Q^2+\omega ^2)]}
     {\ln [(p_t^2+\omega ^2)/\omega ^2]}
\ee
The parameter $\omega$ controls the onset of the suppression. 
For gluons the suppression is stronger than for quarks ($\sim L^2$), 
since they have to be radiated off a parton that is already off-shell. 

\begin{figure}[tbh]\centering
\epsfig{file=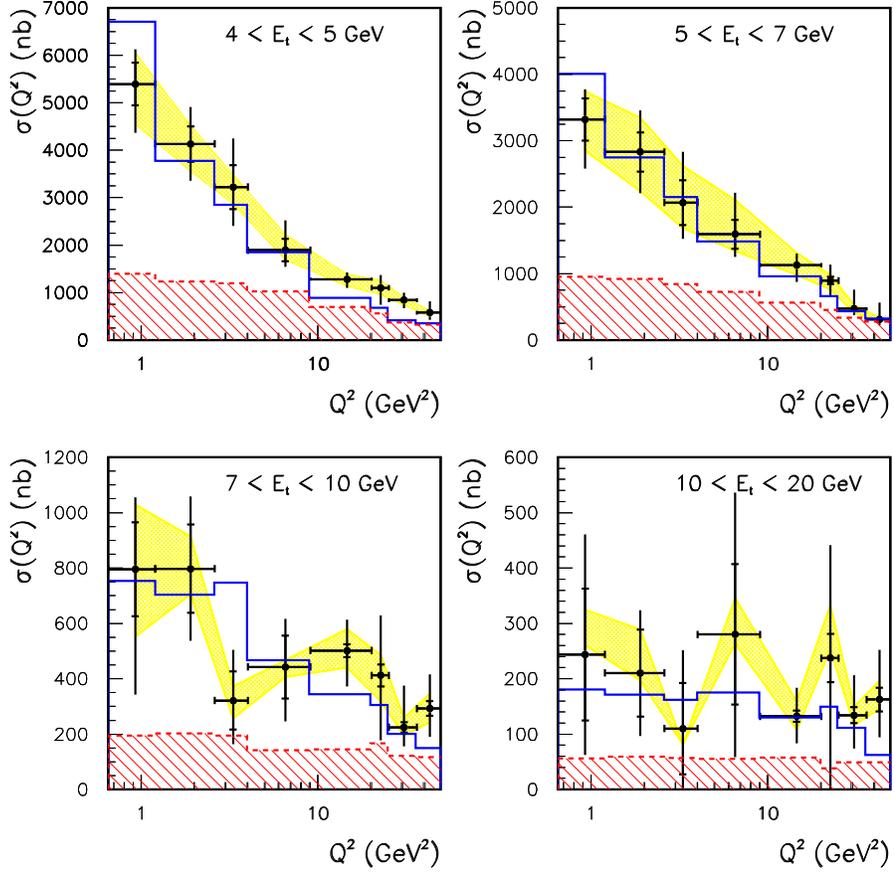,width=13cm}
\vspace{-0.6cm}
\caption[dummy]{\label{fig:virgam}
        $\gamma^*p$ jet cross section as function of $Q^2$
        in ranges of jet transverse energy.
} 
\end{figure}

A study into this direction is the 
measurement of inclusive differential jet cross sections 
at lowest virtualities, in the region $0<Q^2<50$ GeV$^2/c^2$ 
by H1~\cite{h1virgam}. 
The jets are identified here by means of a $k_t$ clustering algorithm
with a resolution scale parameter $E_{cut}=3$ GeV. 
The measurement is expressed as a $\gamma ^*p$ cross section, 
formally defined by dividing the $ep$ cross section by a flux
factor given in the Weizs\"acker-Williams approximation,
thereby correcting for the trivial $Q^2$ dependence due
to the photon propagator. 
The cross section as a function of the virtuality is shown in 
Fig.~\ref{fig:virgam} 
for different regions of transverse jet energy.
The data is compared to predictions obtained with 
the HERWIG Monte Carlo program~\cite{herwig},
which includes in addition to the ``direct'', point-like process
(shown as hatched histogram)
a ``resolved'' contribution. 
The $\gamma ^*$ parton densities
are modeled with the Drees Godbole ansatz and $\omega = 1$GeV$^2/c^2$. 
Only the sum of the two contributions (full histogram) describes the data. 
The ``resolved'' component is found to be important
in the kinematic region where 
the jet transverse energy $E_t^2>Q^2$.

\section{Charm I: $J/\Psi$ Photoproduction}

In the conventional picture of $ep$ interactions, 
charm is produced almost exclusively via 
boson gluon fusion. 
It is therefore considered as the ``classical'' way to probe 
the gluon content of the proton. 
This is one motivation to study the probing process
more closely,
but charm production is an interesting testing ground
of perturbative QCD in its own right.

\subsection{Elastic $J/\Psi$ Photoproduction}

Elastic $J/\Psi$ production offers a very clean experimental signature:
the proton remains intact, 
the vector meson is detected in one of its leptonic decay channels, 
and apart from the $ee$ or $\mu\mu$ pair there is nothing else in the 
detector. 
The energy dependence of the cross section is a characteristic 
feature of the production dynamics. 
In Regge phenomenology, it is related to a Pomeron trajectory 
$
\alpha (t) = 1 + \lambda + \alpha^{\prime}t 
$,
where $t$ is the 4-momentum transfer to the proton~\cite{dola}.   
The cross section behaves like 
\be \label{eq:wdep} 
\sigma \sim \left ( W^2 \right )^{2\lambda} 
\;\; ; 
\ee
$\lambda$ is a universal constant for soft hadronic processes
and of the order 0.08...0.1.
On the other hand, it has been argued that due to the high mass 
of the $J/\Psi$ meson, perturbative QCD should be applicable. 
In the model of Ryskin {\it et al.}~\cite{ryskin},
the dominant diagram contains
a gluon ladder, and the cross section is therefore proportional 
to the square of the gluon density in the proton:
\be
\sigma \sim [xg(x)]^2 
\;\;\; \mbox{\rm with}\;\;\; 
x =  M_{J/\Psi}^2/W^2 \;\; .
\ee
Now the gluon density at low $x$ behaves like $xg(x)\sim x^{-\lambda}$;
for the $W$ dependence again the form 
of Eq.~\ref{eq:wdep}
results, but $\lambda$ has a different meaning.
The rise of the gluon density 
rather gives $\lambda$ values in the region $0.2...0.3$.
Thus, in this model a much steeper rise of the cross section with energy
is expected. 

\begin{figure}[tbh]\centering
\epsfig{file=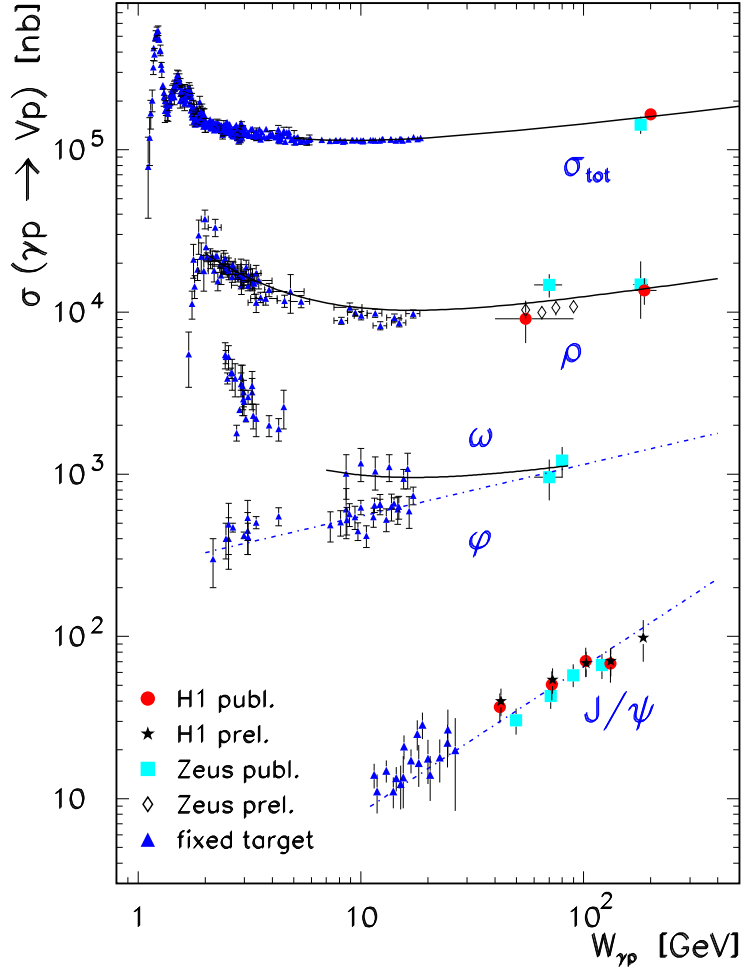,width=10cm}
\vspace{-0.3cm}
\caption[dummy]{\label{fig:vmxsect}
        Total and elastic vector meson photoproduction 
        cross section as function of CMS energy $W_{\gamma p}$.
        The solid lines are parameterizations based on Regge phenomenology.  
} 
\end{figure}

At HERA, the Regge picture is beautifully confirmed by measurements 
of the total photoproduction cross section, 
and by detailed investigations of the 
elastic production of light vector mesons --  
see Fig.~\ref{fig:vmxsect} and 
Ref.~\cite{arndmeyer}
for a recent overview. 
However, the rise of the elastic $J/\Psi$ cross section is found
to be in clear disagreement with a soft Pomeron picture.
This is now seen with the HERA data alone~\cite{jpsi},
for example, ZEUS find
\be
4\lambda = 0.92 \pm  0.14 \pm 0.10 \;\; .
\ee
The $W$ dependence can be well described by the (leading order) 
perturbative QCD calculation in the Ryskin {\it et al.} model, 
the agreement with experimental data is particularly good 
when the MRS(A$^{\prime}$)~\cite{mrs} parameterization of the 
gluon density in the proton is used. 
The sensitivity to the gluon density is temptingly high in 
this model.
However, recently there have also been suggestions to explain
the steep rise of the $J/\Psi$ cross section within a  
modified soft Pomeron framework~\cite{modpom}.

\subsection{Inelastic $J/\Psi$ Photoproduction}

Inelastic $J/\Psi$ production is described in the Color Singlet Model 
(CSM)~\cite{csm} as production of a $\ccb$ pair via boson gluon fusion. 
Another (relatively hard) gluon is radiated off one of the heavy quarks, 
in order to transform the pair into the color-neutral singlet state
forming the $J/\Psi$ meson.
Experimentally, the process can be separated
by means of the elasticity variable   
$z$ that can be calculated 
from the longitudinal momenta of the $J/\psi$ products
and of all final state particles: 
$z = (E-p_z)_{J/\psi}/(E-p_z)_{\mbox{\small all}}$.
It represents, in the proton rest frame, 
the fraction of the photon's energy transferred to the $J/\Psi$ meson. 
Diffractive production leads to $z$ values close to 1, 
``resolved'' contributions can be suppressed by a lower cut on $z$. 

\begin{figure}[tb]\centering
\epsfig{file=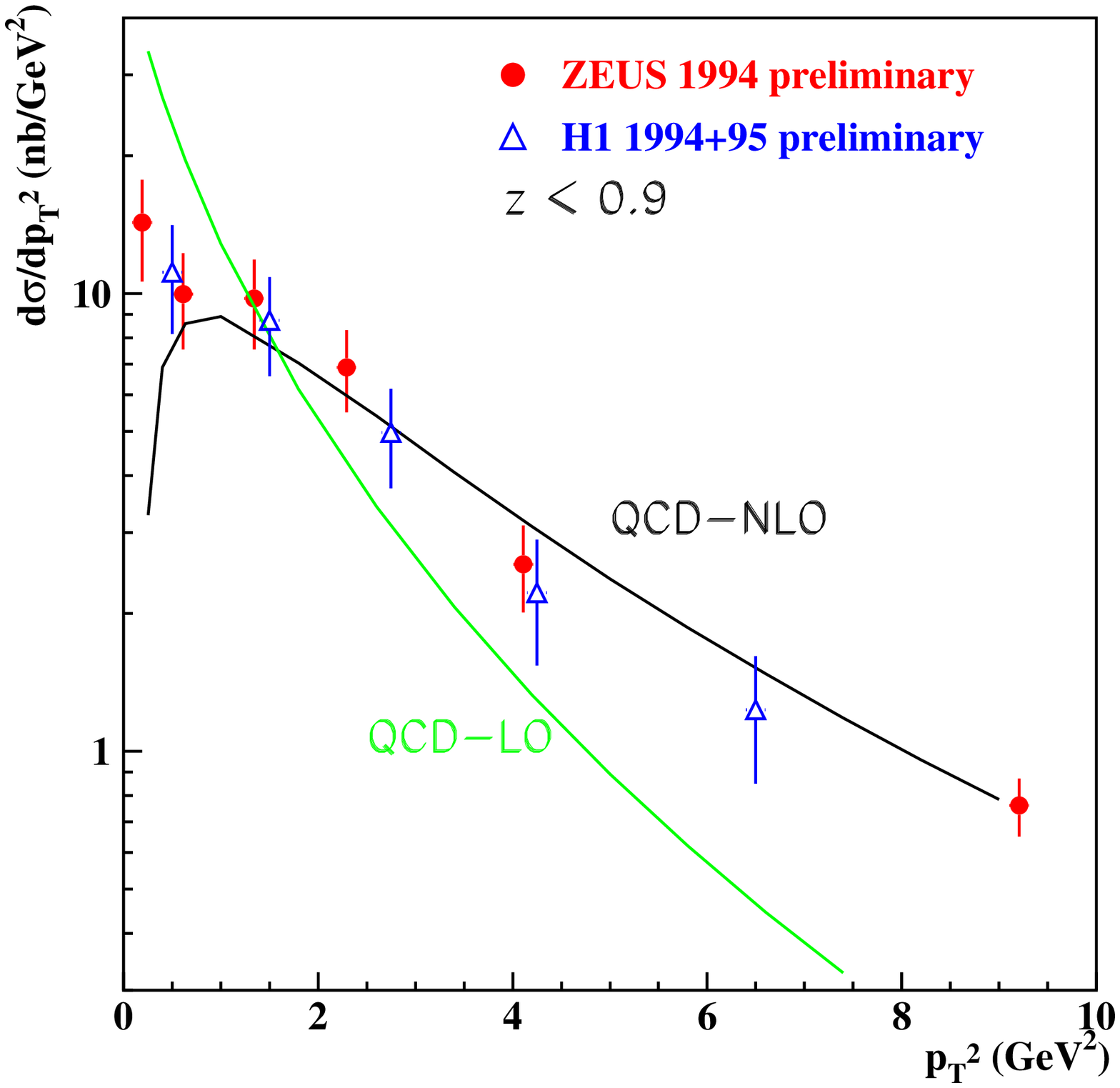,width=5.7cm}
\epsfig{file=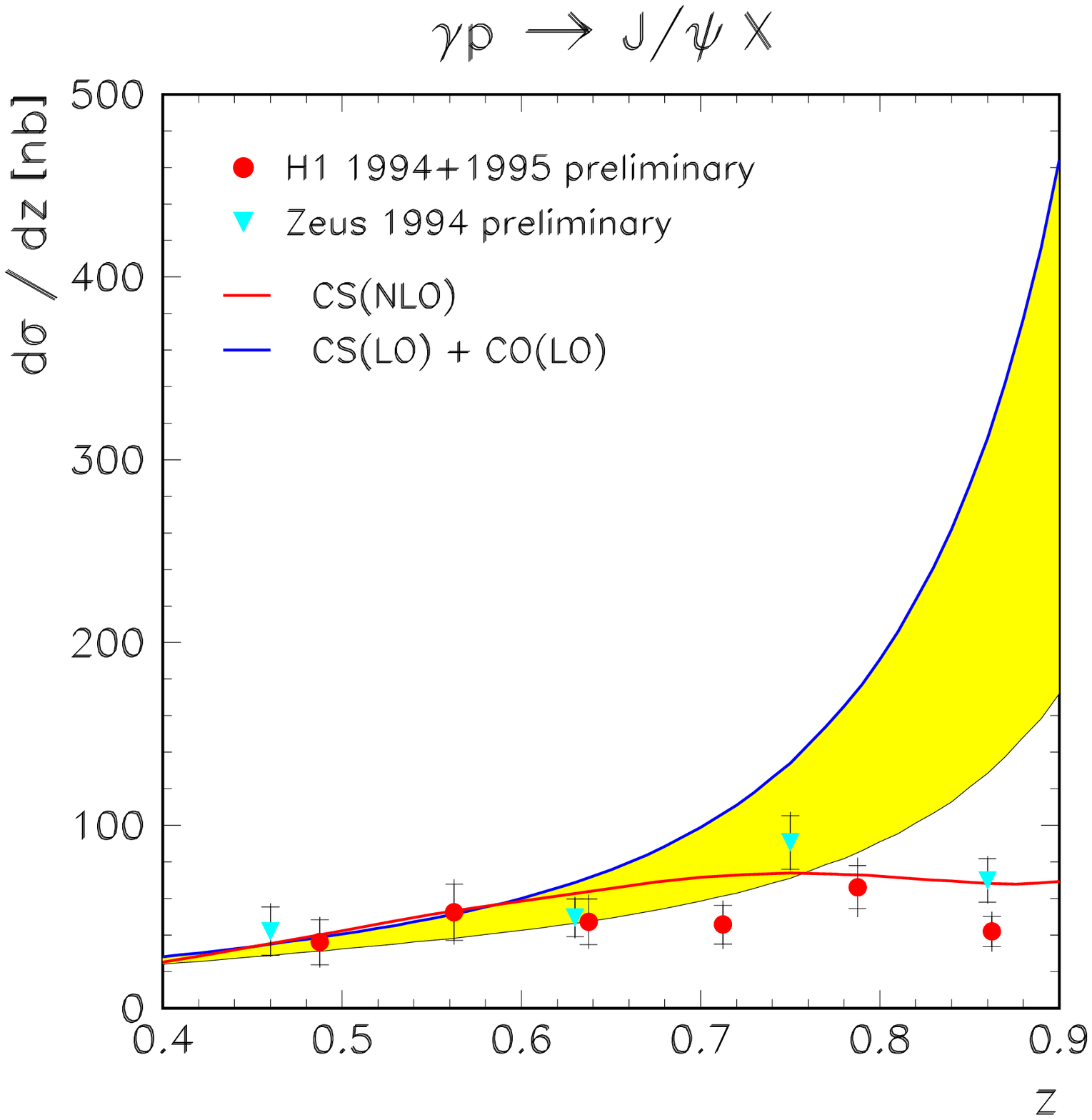,width=5.7cm}

\unitlength1cm
\begin{picture}(0,0)
\put(-1,4.1){(a)}
\put(4.5,4){(b)}
\end{picture}
\vspace{-0.7cm}
\caption[dummy]{\label{fig:psiinel}
          Cross section for inelastic $J/\Psi$ production
          as function of transverse momentum $p_t^2$ (a) and 
          elasticity $z$ (b), compared to QCD calculations in the
          color singlet (CS) model (a)  and CS and color octet (CO) model
          (b). The band indicates the theoretical uncertainty due to 
          initial state radiation.  
} 
\end{figure}

The differential cross section, measured by H1 and ZEUS~\cite{jpsi} 
as function of the $J/\Psi$ transverse
momentum squared is shown in 
Fig.~\ref{fig:psiinel}a 
and compared to QCD calculations~\cite{kraemerpsi} in the CSM, in LO and NLO. 
Next to Leading Order corrections are substantial and 
clearly needed to reproduce the data. 
The parameters in the theoretical calculation 
(charm mass $m_c = 1.4$ GeV/$c^2$, 
renormalization and factorization scale $\mu = \sqrt{2}m_c$, 
and $\Lambda_{QCD}=200$ MeV) 
are slightly ``stretched''; other choices 
tend to give a lower cross section normalization 
and would leave room for additional production mechanisms.
e.g.\  via fragmentation~\cite{kniehlpsifrag} or 
through color octet states. 

Recently, the r\^ole of color octet contributions has been intensively 
discussed, as a possible explanation for the unexpectedly high 
$J/\Psi$ and $\Psi^{\prime}$ production rates measured at the 
Tevatron~\cite{psitev}.
In the factorization approach of non-relativistic QCD~\cite{nrqcd}, 
$J/\Psi$ mesons can also be produced via color-octet $\ccb$
configurations. 
The transition to the singlet final state
is treated as a soft process described by 
non-perturbative matrix elements,
which must be obtained from experimental data. 
They have been extracted~\cite{leibovich} from the CDF $J/\Psi$ rates 
and yield a prediction for the HERA regime that can be confronted to data. 
The cross section measured by H1 and ZEUS is shown as a function of 
the elasticity $z$ in Fig.~\ref{fig:psiinel}b.
The color octet contributions are expected to enhance the cross section
in the high $z$ region~\cite{coathera}.  
No indication for such an enhancement can be seen in the data yet.
However, there are considerable theoretical uncertainties in the 
prediction. 
As an example, the band in the figure indicates the range of uncertainty
due to an effective transverse momentum of the incoming partons, 
due to initial state radiation~\cite{octetkt}. 


\section{Charm II: $D^*$ Production}

Open charm at HERA has mainly been reconstructed in the 
fragmentation and decay chain 
$c \ra D^{*+} \ra D^0\pi^+ \ra (K^-\pi^+)\pi^+$
exploiting the clean experimental signal in the mass difference 
distribution of $m (K^-\pi^+)-m(K^-\pi^+)$.
 
\subsection{$D^*$ Production by Real Photons}

\begin{figure}[tbh]\centering
\epsfig{file=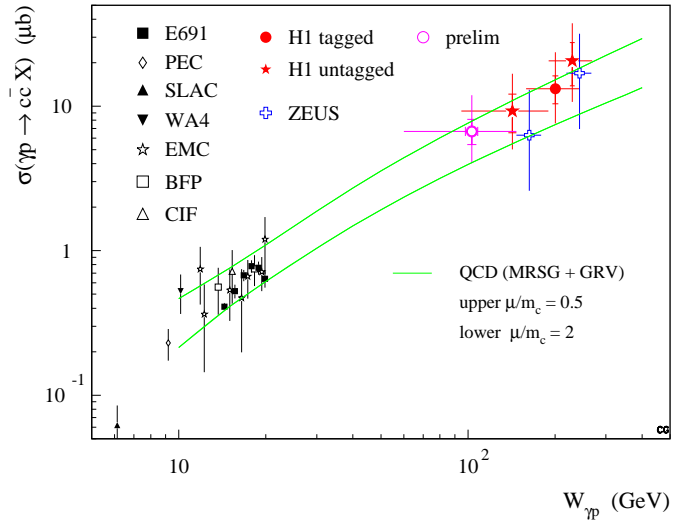,width=9cm}
\vspace{-0.5cm}
\caption[dummy]{\label{fig:dstarstot}
        Photoproduction cross section as function of $Q^2$
        in ranges of jet transverse energy.
} 
\end{figure}

The total open charm photoproduction cross section
has been measured at HERA~\cite{dstargp} 
at different photon proton center-of-mass energies
up to $W\approx 230$ GeV
and found to be an order of magnitude higher
than at fixed target energies
(Fig.~\ref{fig:dstarstot}).
Together with a new point at ``low'' $W\approx 90$ GeV 
the data start to provide some lever arm on the slope
of the cross section versus energy, which is related to 
the $x$ dependence of the gluon density in the proton. 
However, there are large factors for the extrapolation (in $\eta$ and 
mainly $p_t$) from the visible to the total cross section involved,
which depend on the parton densities in the proton and the photon.
It is thus not straightforward to interpret the total cross section data 
in these terms.    

Currently there exist two approaches to calculate 
differential distributions for $D^*$ photoproduction 
in NLO QCD. 
In the so-called ``massive'' charm scheme~\cite{frixione},
charm quarks are treated as massive particles 
which are only generated perturbatively in the final state, 
whereas the 3 lightest flavours (and gluons) are the only active 
partons in the initial state proton and photon. 
This scheme is valid as long as $p_t$ is not too large 
with respect to the charm mass,
and indispensable for the calculation of the total cross section.
In contrast, in the ``massless'' scheme~\cite{kniehldstar,cacciaridstar},
charm is also an active flavour in the parton distributions 
of the proton and the photon. 
This scheme should work particularly well for large $p_t\gg m_c$;
here, scale-dependent perturbative fragmentation functions 
are needed to control divergences.
The ``massless'' scheme predicts a large ($>50\%$) contribution 
of ``resolved'' processes of the type $cg \ra cg$ initiated by a $c$
quark from the photon.%
\footnote{This has no equivalent in the ``massive'' scheme, but one 
should bear in mind that only the sum of ``direct'' and ``resolved'' 
contributions is unambiguously defined at NLO.}
For comparison, the ``resolved'' component due to
due to $gg \ra\ccb$, 
present in both 
approaches, is only of order 10\%.    

\begin{figure}[tbh]\centering
\epsfig{file=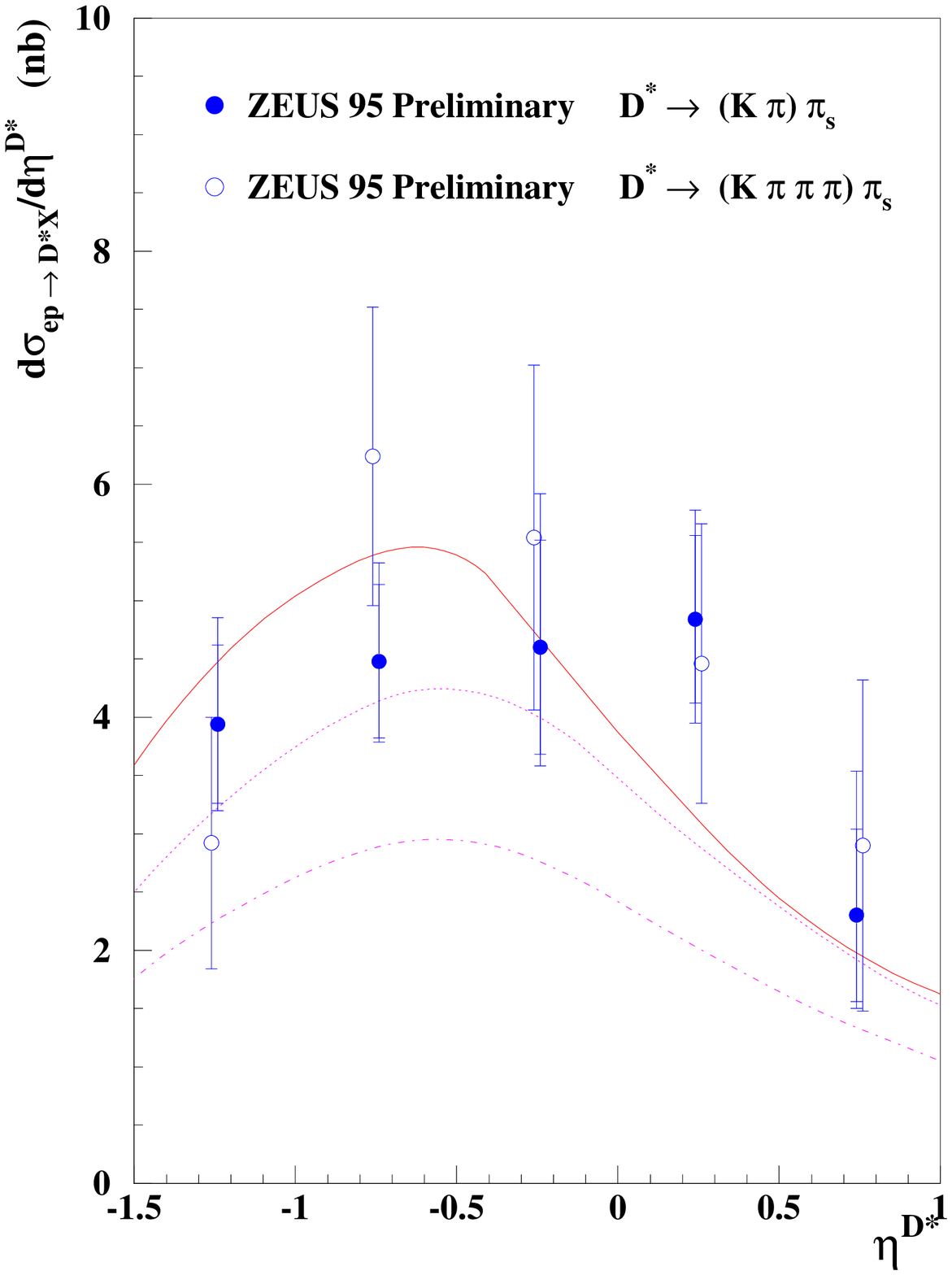,width=5cm}
\epsfig{file=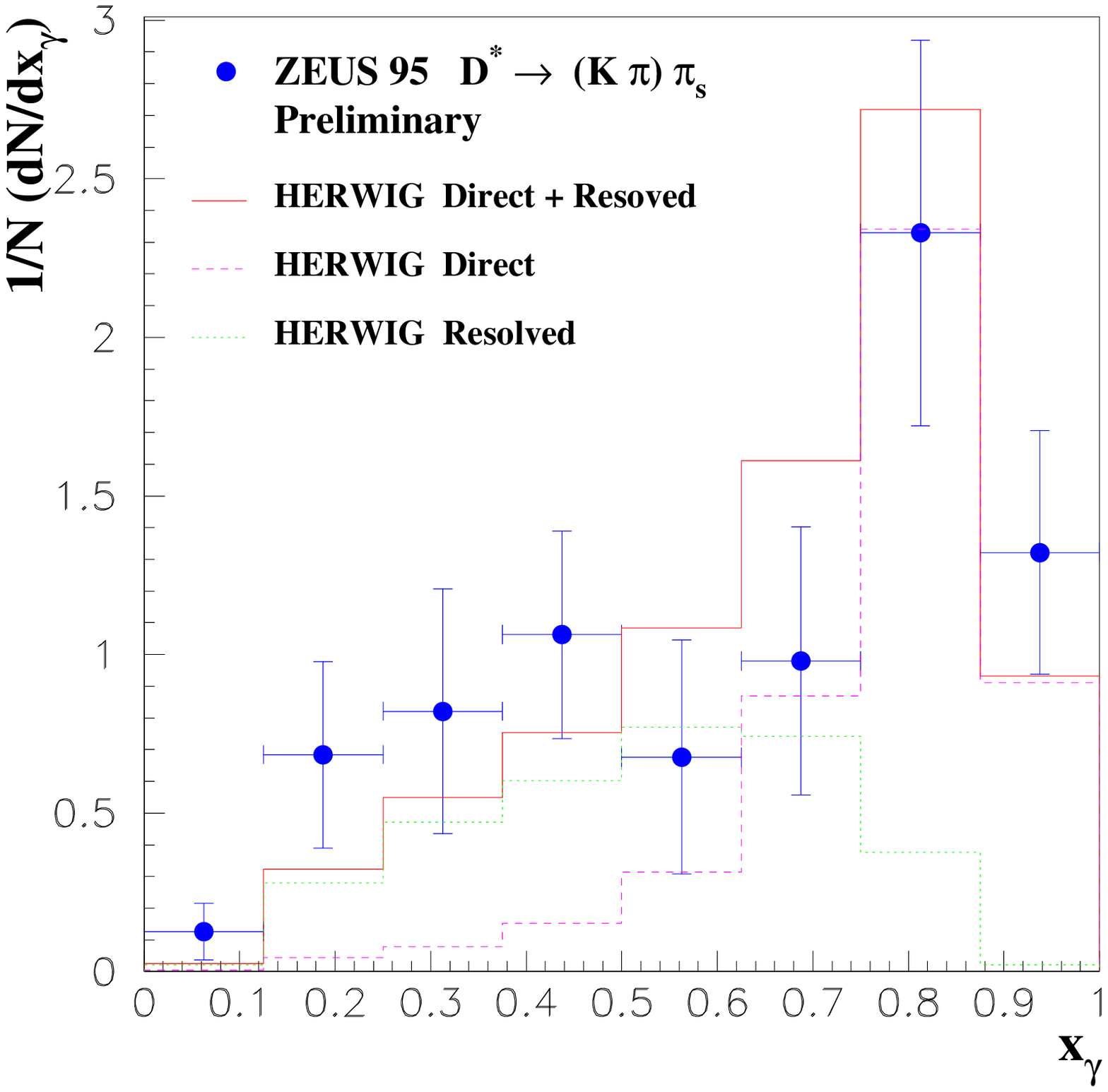,width=6.8cm}

\unitlength1cm
\begin{picture}(0,0)
\put(-2,4){(a)}
\put(0.4,4){(b)}
\end{picture}
\vspace{-1cm}
\caption[dummy]{\label{fig:dstgp}
          (a) Differential cross section for $D^*$ photoproduction
              as function of rapidity.
              The curves represent NLO QCD calculations in 
              the ``massless'' (upper) and ``massive'' scheme,
              with default parameters (lower), and with 
              charm mass $m_c=$ 1.5 GeV$/c^2$ and scale $\mu=0.5m_c$
              (middle).  
          (b) Uncorrected $x_{\gamma}^{OBS}$ distribution 
              for dijet events  containing a $D^*$ meson. 
} 
\end{figure}

Differential cross sections as function of (pseudo-)rapidity and 
$p_t$ have been measured by H1
and ZEUS~\cite{dstargp},  
with lower $p_t$ cuts at 2.5 or 3 GeV$/c$.
The bulk of the data thus lies in a region where $p_t$ 
is neither particularly ``small'' nor ``large''.
In Fig.~\ref{fig:dstgp}a), the ZEUS results are compared 
to NLO QCD calculations in the two schemes. 
Within the -- still large -- experimental errors, 
the shape of the distribution is reproduced in both approaches. 
The normalization prefers the ``massless' calculation,
which has been performed using
the Peterson-Zerwas form of the fragmentation function~\cite{peterson} with
a fragmentation parameter 
$\epsilon _c = 0.064$  
as it was obtained from fits to $e^+e^-$ data within the same
framework. 
Within the ``massive'' scheme, such an analysis has not yet been 
performed; the parameter $\epsilon _c = 0.060$ 
is taken from a leading order analysis~\cite{chrin} of low
energy $e^+e^-$ data. 
It has been argued~\cite{cacciaridstar} that rather a smaller value
$\epsilon _c \approx 0.020$ would be consistent with the
NLO calculation and reconcile it with the data. 

In order to obtain further information on the production process, 
ZEUS has performed a dijet analysis similar to that 
described in Sec.~\ref{sec:dijetgp}
and measured the uncorrected $x_{\gamma}^{OBS}$ distribution for events 
containing a $D^*$ meson~\cite{dstargpxgam}.
The jets were identified using a cone algorithm or -- shown here --
a $k_t$ cluster algorithm and were required to have 
transverse energy above 4 GeV.    
The result is shown in Fig.~\ref{fig:dstgp}b 
and compared to the predictions of the HERWIG leading order 
Monte Carlo program.
The data at lower $x_{\gamma}^{OBS}$ values
need a substantial ``resolved'' photon contribution, 
which in the Monte Carlo is dominated
by processes initiated by charm quarks from the photon.

\subsection{$D^*$ Production by Virtual Photons}

In deep inelastic scattering, due to the recoil of the scattered 
electron, $D^*$ production can be measured in the full range down to
$p_t=0$ in the hadronic center-of-mass system, although a lower 
cut on the transverse momentum in the laboratory frame is usually
necessary to suppress combinatorial background. 
The data is then dominated by $p_t\approx m_c$, so that one 
can expect a calculation in the ``massive'' 
scheme with 
3 active flavours 
to provide a good description of the process 
in this regime. 
\begin{figure}[tbh]\centering
\epsfig{file=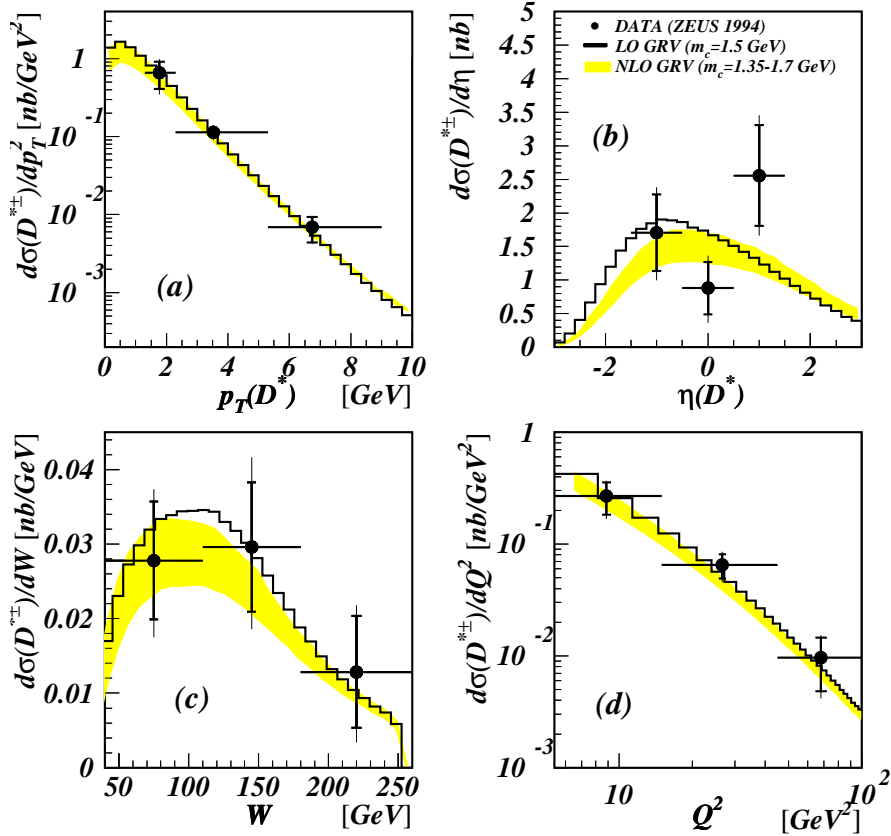,width=12cm}
\vspace{-0.6cm}
\caption[dummy]{\label{fig:dstardis}
        Differential cross sections for deep inelastic $D^*$ production.       
} 
\end{figure}
Fig.~\ref{fig:dstardis}
demonstrates that this is indeed the case:
the NLO calculation by Harris and Smith~\cite{harrissmith}, 
performed in this scheme, 
agrees well with the data~\cite{dstardis}
in both magnitude and shape. 
One should also note that due to the smallness of the 
charm transverse momenta, 
$p_t^2 \ll Q^2$ is in general fulfilled, 
so that the virtual photon can be considered as a point-like probe to 
a good approximation. 
Deep inelastic charm production is therefore a very promising
process to extract information about the gluon content of 
the proton. 

\section{Conclusion}

Jets and heavy quarks carry the information 
from the partonic interactions in $ep$ collisions. 
The new HERA results in this field
as discussed here can be successfully described
in the framework of perturbative QCD.
Due to the large kinematic range at HERA, complementary 
concepts have to be applied. Charm quarks appear as massive 
products of boson gluon fusion
-- or as massless partons 
in the proton and even the photon. 
Virtual photons can be considered as point-like probes 
of the hadron structure, but they also reveal their nature
as composite hadronic objects as it is well explored now for 
their real counterpart. 
It is this variety in the HERA phenomenology that holds the promise 
of providing keys to new dimensions in our understanding
of the strong interaction.  

\section*{Acknowledgments}
It is a pleasure to thank my colleagues in the H1 and ZEUS collaboration
for providing the results for this review.
I benefitted from conversations with B.~Kniehl. 
I would like to thank B.~Foster and his team 
for creating the stimulating atmosphere of the conference, 
and for the warm welcome I enjoyed at Bristol.


\section*{References}
\begin{thebibliography}{99}
\bibitem{zeusjetsh}
      M. Mart\'\i nez, 
      Jet Shapes at HERA, 
      talk at ``DIS 97", Chicago, 1997, to appear in the proceedings. 
\bibitem{pythia}
      T.\ Sj\"ostrand, CERN-TH-6488 (1992), 
      Comp.\,Phys.\,Comm.\ 82 (1994) 74.
\bibitem{jetsh_LEP}
      R. Akers {\it et al.} (OPAL Coll.),
      Z. Phys.\, C 63 (1994) 197.
\bibitem{jetsh_pp}
      F. Abe {\it et al.} (CDF Coll.),
      Phys.\,Rev.\,Lett.\ 70 (1993) 713, \\
      S. Abachi {\it et al.} (D0 Coll.),
      Phys.\,Lett.\ B357 (1995) 500.
\bibitem{klasenjetsh}
      M. Klasen, G. Kramer, 
      Phys.\,Rev.\,D56 (1997) 2702.
\bibitem{zeusdijetang}
      M. Derrick {\it et al.} (ZEUS Coll.),
      Phys.\,Lett.\,B 384 (1996) 401.
\bibitem{harrisowensjets}
      B.W. Harris, J.F. Owens,
      Phys.\,Rev.\,D56 (1997) 4007.
\bibitem{h1gxgamma}
      T.\,Ahmed {\it et al.} (H1 Coll.).
      N.\,Phys.\ B445 (1995) 195.
\bibitem{h1fxgeff}
      C. Adloff  {\it et al.} (H1 Coll.),
      preprint DESY 97-164.
\bibitem{combmaxwell} 
      B.L.Combridge, C.J.Maxwell, 
      Nucl.\,Phys.\ B239 (1984) 429.
\bibitem{grv_gamma}
      M. Gl\"uck, E. Reya, A. Vogt, 
      Phys.\,Rev.\,D46 (1992) 1973.
\bibitem{rickphd}
      H. Rick (H1 Coll.), 
      PhD thesis, Universit\"at Dortmund, Germany, 1997. 
\bibitem{f2gammacompi}
      Ch.\,Berger {\it et al.} (PLUTO Coll.),
      Nucl.\ Phys.\ B281 (1987) 365; \\
      W.\,Bartel {\it et al.} (JADE Coll.), 
      Z.\ Phys.\  C24 (1984) 231; \\
      M.\,Althoff {\it et al.} ( TASSO Coll.), 
      Z.\ Phys.\  C31 (1986) 527;\\
      H.\,Aihara {\it et al.} (TPC/Two-Gamma Coll.), 
      Z.\ Phys.\  C34 (1987) 1;\\
      S.K.\,Sahu {\it et al.} (AMY Coll.), 
      Phys.\ Lett.\  B346 (1995) 208;\\
      K.\,Muramatsu {\it et al.} (TOPAZ Coll.), 
      Z.\ Phys.\  C31 (1986) 527;\\
      P.\,Abreu {\it et al.} (DELPHI Coll.), 
      Z.\ Phys.\  C69 (1996) 223;\\
      R.\,Akers {\it et al.} (OPAL Coll.), 
      CERN-PPE/93-156, August 1993.
\bibitem{fkp}
      J.H.\,Field, F.\,Kapusta, L.\,Poggioli,
      Z.\ Phys.\ {\bf C36} (1987) 121
\bibitem{spiekermoriond}
      J. Spiekermann,
      talk at ``32nd Rencontres de Moriond: QCD and 
      high-energy 
      hadronic interactions'', Les Arcs, France, 1997, 
      to appear in the proceedings. 
\bibitem{ariadne}
      L. L\"onnblad,
      Comp.\,Phys.\,Comm.\ 71 (1992) 15.
\bibitem{lepto}
      G.Ingelman,
      in: Proc. ``Physics at HERA'', Hamburg, 1991, 
      eds.\, W. Buchm\"uller, G. Ingelman, Vol.\, 3;\\
      G. Ingelman, A. Edin, J. Rathsman, 
      Comp.\,Phys.\,Comm.\ 101 (1997) 108.
\bibitem{disent}
      S. Catani, M. Seymour, 
      Nucl.\,Phys.\ B485 (1997) 291.
\bibitem{sasgam}
      G. Schuler, T. Sj\"ostrand, 
      Z.\ Phys.\  C68 (1995) 607, 
      Phys.\,Lett.\,B 376 (1996) 193;\\
      M. Gl\"uck, E.\ Reya, M. Stratman, 
      Phys.\,Rev.\,D54 (1996) 5515.
\bibitem{dreesgodbole}
      M. Drees, R. Godbole, 
      Phys.\,Rev.\,D50 (1994) 3124.
\bibitem{borzumati}
      F. Borzumati, G. Schuler,
      Z.\ Phys.\  C58 (1993) 139. 
\bibitem{h1virgam}
      C. Adloff  {\it et al.} (H1 Collaboration),
      preprint DESY 97-164.
\bibitem{herwig}
      G. Marchesini et al.,
      Comp.\,Phys.\,Comm.\ 67 (1992) 465.
\bibitem{dola}
      A.\ Donnachie, P.V.\ Landshoff, 
      Phys.\,Lett.\  B348 (1995) 213, 
      ibid.\ B296 (1992) 227.
\bibitem{ryskin} 
      M.G.\ Ryskin, 
      Z.\,Phys.\ C57 (1993) 89; \\ 
      M.G.\ Ryskin, R.G.\ Roberts, A.D.\ Martin, E.M.\ Levin
      preprint DTP/95/96 CBPF-NF-079/95, RAL-TR-95-065, 
      revised February 1996.
\bibitem{arndmeyer}
      A. Meyer, 
       talk at "5th Topical Seminar 
      on the Irresistible Rise of the Standard Model",
      San Miniato, Italy, 1997,
      to appear in the proceedings. 
\bibitem{jpsi}
      S.\,Aid et al.\ (H1 Coll.), 
      Nucl.\,Phys.\, B472 (1996) 3, 
      and 
      ``Diffractive and non-diffractive photoproduction
      of $J/\Psi$ mesons at HERA'', 
      contr.\ paper to ICHEP 1996, Warsaw; \\ 
      J. Breitweg {\it et al.} (ZEUS Coll.),
      Z.\ Phys.\  C75 (1997) 215,
      and 
      preprint DESY 97-147.
\bibitem{mrs}
      A.D.\ Martin, R.G.\ Roberts, W.J.\ Stirling,    
      Phys.\,Lett.\ B354 (1995) 155. 
\bibitem{modpom}
      L.P.A. Haakman, A.Kaidalov, J.H. Koch, 
      Phys.\,Lett.\  B365 (1996) 411; 
      L.L. Jenjovszky, E.S. Martynov, F. Paccanoni, 
      Novy Svet Hadrons (1996) 170. 
\bibitem{csm}
      E.L.\ Berger, D.\ Jones,
      Phys.\, Rev.\ D23 (1981) 1521, \\
      R. Baier, R. R\"uckl,
      Nucl.\,Phys.\ B201 (1982) 1. 
\bibitem{kraemerpsi}
      M.\ Kr\"amer, J.\ Zunft, J.\ Steegborn, P.M.\ Zerwas, 
      Phys.\, Lett.\ B348 (1995) 657;
      M.\ Kr\"amer, 
      Nucl.\ Phys.\ B459 (1996) 3.     
\bibitem{kniehlpsifrag} 
      B.A. Kniehl, G. Kramer, 
      preprints DESY-97-036, DESY-97-110.
\bibitem{psitev}
      F.\ Abe et al.\ (CDF Coll.),
      Phys.\,Rev.\,Let.\ 69 (1992) 3704; \\
      A.\ Sansoni et al.\ (CDF Coll.),
      Nuovo Cim.\, 109A (1996) 827.
\bibitem{nrqcd}
      G.T.\ Bodwin, E.\ Braaten, G.P.\ Lepage, 
      Phys.\,Rev.\ D51 (1995) 1125. 
\bibitem{leibovich}
      P. Cho, A.K. Leibovich,
      Phys.\, Rev.\, D53 (1996) 150, Phys.\,Rev.\, D53 (1996) 6203.
\bibitem{coathera}
      M.\ Cacciari, M.\ Kr\"amer, 
      Phys.\,Rev.\,Lett.76 (1996) 4128.
\bibitem{octetkt}
      B. Cano-Coloma, M.A. Sanchis-Lozano,
      Valencia Univ.\ preprint IFIC-97-29, hep-ph/9706270.
\bibitem{dstargp} 
      M.\ Derrick et al.\ (ZEUS Coll.), 
      Phys.\,Lett.\ B349 (1995) 225;\\
      S. Aid {\it et al.} (H1 Coll.),
      Nucl.\,Phys.\ B472 (1996) 32.
\bibitem{dstargpxgam}
      C. Coldewey, 
      talk at Ringberg Workshop on new trends in HERA physics,
      Ringberg, Germany, 1997, to appear in the proceedings.
\bibitem{frixione} 
      S.\ Frixione et al., 
      Phys.\,Lett.\ B348 (1995) 633, 
      Nucl.\,Phys.\ B454~(1995)~3.
\bibitem{kniehldstar}
      B.A. Kniehl, G. Kramer, M. Spira,
      preprint DESY-96-210;\\
      J. Binnewies, B.A. Kniehl, G. Kramer,
      preprint DESY-97-012.
\bibitem{cacciaridstar} 
      M. Cacciari, M. Greco, 
      Phys.\, Rev.\, D55 (1997) 7134.
\bibitem{peterson} 
      C. Peterson, D. Schlatter, I. Schmitt, P.M. Zerwas, 
      Phys.\, Rev.\ D 27 (1983) 105.
\bibitem{chrin} 
      J. Chrin, 
      Z.\,Phys.\ C 36 (1987) 163.
\bibitem{harrissmith}
      B.W. Harris, J. Smith, 
      Nucl.\,Phys.\ B452 (1995) 109;       
      Phys.\,Lett.\ B353 (1995) 535;
      preprint DESY 97-111.           
\bibitem{dstardis} 
      C. Adloff  {\it et al.} (H1 Collaboration),
      Z.\,Phys.\  C72 (1996) 593; \\ 
      J. Breitweg {\it et al.} (ZEUS Coll.),
      preprint DESY 97-089.

\end{thebibliography}
\end{document}